\documentstyle[twoside]{article}

\catcode`\@=11
\long\def\@makefntext#1{
\protect\noindent \hbox to 3.2pt {\hskip-.9pt  
$^{{\eightrm\@thefnmark}}$\hfil}#1\hfill}		

\def\@makefnmark{\hbox to 0pt{$^{\@thefnmark}$\hss}}	
	
\def\ps@myheadings{\let\@mkboth\@gobbletwo
\def\@oddhead{\hbox{}
\rightmark\hfil\eightrm\thepage}   
\def\@oddfoot{}\def\@evenhead{\eightrm\thepage\hfil
\leftmark\hbox{}}\def\@evenfoot{}
\def\sectionmark##1{}\def\subsectionmark##1{}}

\oddsidemargin=\evensidemargin
\newcounter{sectionc}\newcounter{subsectionc}\newcounter{subsubsectionc}
\renewcommand{\section}[1] {\vspace{12pt}\addtocounter{sectionc}{1} 
\setcounter{subsectionc}{0}\setcounter{subsubsectionc}{0}\noindent 
	{\tenbf\thesectionc. #1}\par\vspace{5pt}}
\renewcommand{\subsection}[1] {\vspace{12pt}\addtocounter{subsectionc}{1} 
	\setcounter{subsubsectionc}{0}\noindent 
	{\bf\thesectionc.\thesubsectionc. {\kern1pt \bfit #1}}\par\vspace{5pt}}
\renewcommand{\subsubsection}[1] {\vspace{12pt}\addtocounter{subsubsectionc}{1}
	\noindent{\tenrm\thesectionc.\thesubsectionc.\thesubsubsectionc.
	{\kern1pt \tenit #1}}\par\vspace{5pt}}
\newcommand{\nonumsection}[1] {\vspace{12pt}\noindent{\tenbf #1}
	\par\vspace{5pt}}

\newcounter{appendixc}
\newcounter{subappendixc}[appendixc]
\newcounter{subsubappendixc}[subappendixc]
\renewcommand{\thesubappendixc}{\Alph{appendixc}.\arabic{subappendixc}}
\renewcommand{\thesubsubappendixc}
	{\Alph{appendixc}.\arabic{subappendixc}.\arabic{subsubappendixc}}

\renewcommand{\appendix}[1] {\vspace{12pt}
        \refstepcounter{appendixc}
        \setcounter{figure}{0}
        \setcounter{table}{0}
        \setcounter{lemma}{0}
        \setcounter{theorem}{0}
        \setcounter{corollary}{0}
        \setcounter{definition}{0}
        \setcounter{equation}{0}
        \renewcommand{\thefigure}{\Alph{appendixc}.\arabic{figure}}
        \renewcommand{\thetable}{\Alph{appendixc}.\arabic{table}}
        \renewcommand{\theappendixc}{\Alph{appendixc}}
        \renewcommand{\thelemma}{\Alph{appendixc}.\arabic{lemma}}
        \renewcommand{\thetheorem}{\Alph{appendixc}.\arabic{theorem}}
        \renewcommand{\thedefinition}{\Alph{appendixc}.\arabic{definition}}
        \renewcommand{\thecorollary}{\Alph{appendixc}.\arabic{corollary}}
        \renewcommand{\theequation}{\Alph{appendixc}.\arabic{equation}}
        \noindent{\tenbf Appendix \theappendixc #1}\par\vspace{5pt}}
\newcommand{\subappendix}[1] {\vspace{12pt}
        \refstepcounter{subappendixc}
        \noindent{\bf Appendix \thesubappendixc. {\kern1pt \bfit #1}}
	\par\vspace{5pt}}
\newcommand{\subsubappendix}[1] {\vspace{12pt}
        \refstepcounter{subsubappendixc}
        \noindent{\rm Appendix \thesubsubappendixc. {\kern1pt \tenit #1}}
	\par\vspace{5pt}}

\newcommand{\textlineskip}{\baselineskip=13pt}
\newcommand{\smalllineskip}{\baselineskip=10pt}

\def\eightcirc{
\begin{picture}(0,0)
\put(4.4,1.8){\circle{6.5}}
\end{picture}}
\def\eightcopyright{\eightcirc\kern2.7pt\hbox{\eightrm c}} 

\newcommand{\copyrightheading}[1]
	{\vspace*{-2.5cm}\smalllineskip{\flushleft
	{\footnotesize International Journal of Modern Physics B, #1}\\
	{\footnotesize $\eightcopyright$\, World Scientific Publishing
	 Company}\\
	 }}

\newcommand{\publisher}[2]{{\begin{center}\footnotesize\smalllineskip 
	Received #1\\
	Revised #2
	\end{center}
	}}

\def\abstracts#1#2#3{{
	\centering{\begin{minipage}{4.5in}\baselineskip=10pt\footnotesize
	\parindent=0pt #1\par 
	\parindent=15pt #2\par
	\parindent=15pt #3
	\end{minipage}}\par}} 

\def\keywords#1{{
	\centering{\begin{minipage}{4.5in}\baselineskip=10pt\footnotesize
	{\footnotesize\it Keywords}\/: #1
	\end{minipage}}\par}}

\renewenvironment{thebibliography}[1]			
	{\frenchspacing
	 \ninerm\baselineskip=11pt
	 \begin{list}{\arabic{enumi}.}
	{\usecounter{enumi}\setlength{\parsep}{0pt}
	 \setlength{\leftmargin 12.7pt}{\rightmargin 0pt} 
	 \setlength{\itemsep}{0pt} \settowidth
	{\labelwidth}{#1.}\sloppy}}{\end{list}}

\newcounter{itemlistc}
\newcounter{romanlistc}
\newcounter{alphlistc}
\newcounter{arabiclistc}

\newcommand{\fcaption}[1]{
        \refstepcounter{figure}
        \setbox\@tempboxa = \hbox{\footnotesize Fig.~\thefigure. #1}
        \ifdim \wd\@tempboxa > 5in
           {\begin{center}
        \parbox{5in}{\footnotesize\smalllineskip Fig.~\thefigure. #1}
            \end{center}}
        \else
             {\begin{center}
             {\footnotesize Fig.~\thefigure. #1}
              \end{center}}
        \fi}

\newcommand{\tcaption}[1]{
        \refstepcounter{table}
        \setbox\@tempboxa = \hbox{\footnotesize Table~\thetable. #1}
        \ifdim \wd\@tempboxa > 5in
           {\begin{center}
        \parbox{5in}{\footnotesize\smalllineskip Table~\thetable. #1}
            \end{center}}
        \else
             {\begin{center}
             {\footnotesize Table~\thetable. #1}
              \end{center}}
        \fi}

\def\@citex[#1]#2{\if@filesw\immediate\write\@auxout
	{\string\citation{#2}}\fi
\def\@citea{}\@cite{\@for\@citeb:=#2\do
	{\@citea\def\@citea{,}\@ifundefined
	{b@\@citeb}{{\bf ?}\@warning
	{Citation `\@citeb' on page \thepage \space undefined}}
	{\csname b@\@citeb\endcsname}}}{#1}}

\newif\if@cghi
\def\cite{\@cghitrue\@ifnextchar [{\@tempswatrue
	\@citex}{\@tempswafalse\@citex[]}}
\def\citelow{\@cghifalse\@ifnextchar [{\@tempswatrue
	\@citex}{\@tempswafalse\@citex[]}}
\def\@cite#1#2{{$\null^{#1}$\if@tempswa\typeout
	{IJCGA warning: optional citation argument 
	ignored: `#2'} \fi}}

\def\pmb#1{\setbox0=\hbox{#1}
	\kern-.025em\copy0\kern-\wd0
	\kern.05em\copy0\kern-\wd0
	\kern-.025em\raise.0433em\box0}

\def\fnt#1#2{\footnotetext{\kern-.3em
	{$^{\mbox{\scriptsize #1}}$}{#2}}}

\def\fpage#1{\begingroup
\voffset=.3in
\thispagestyle{empty}\begin{table}[b]\centerline{\footnotesize #1}
	\end{table}\endgroup}

\font\tenrm=cmr10
\font\tenit=cmti10 
\font\tenbf=cmbx10
\font\bfit=cmbxti10 at 10pt
\font\ninerm=cmr9

\font\eightrm=cmr8

\def\qed{\hbox{${\vcenter{\vbox{			
   \hrule height 0.4pt\hbox{\vrule width 0.4pt height 6pt
   \kern5pt\vrule width 0.4pt}\hrule height 0.4pt}}}$}}

\def\bsc{{\sc a\kern-6.4pt\sc a\kern-6.4pt\sc a}}	
\def\bflatex{\bf L\kern-.30em\raise.3ex\hbox{\bsc}\kern-.14em 
T\kern-.1667em\lower.7ex\hbox{E}\kern-.125em X} 

\begin{document}


\normalsize\textlineskip
\thispagestyle{empty}
\setcounter{page}{1}

\copyrightheading{}			

\vspace*{0.88truein}

\fpage{1}
\centerline{\bf ON THE NUMBER OF METASTABLE STATES IN A STRIPE GLASS}
\vspace*{0.035truein}
\vspace*{0.37truein}
\centerline{\footnotesize J\"ORG SCHMALIAN and  HARRY WESTFAHL JR.} 
\vspace*{0.015truein}
\centerline{\footnotesize\it Department of Physics and Astronomy and Ames Laboratory, Iowa 
State University} 
\baselineskip=10pt
\centerline{\footnotesize\it Ames, Iowa 50011, USA} 
\vspace*{10pt}
\centerline{\normalsize and}
\vspace*{10pt}
\centerline{\footnotesize PETER G. WOLYNES}
\vspace*{0.015truein}
\centerline{\footnotesize\it Department of Chemistry and Biochemistry, University of California, San Diego}
\baselineskip=10pt
\centerline{\footnotesize\it La Jolla, California 92093, USA}
\vspace*{0.225truein}
\publisher{(received date)}{(revised date)}

\vspace*{0.21truein}
\abstracts{We estimate the number of metastable states of a self generated stripe glass,
 relevant for the formation of glassy doped Mott insulators. Using replica bound states, we
demonstrate that  the configurational entropy  is the difference between  
the entropy of the stripe liquid  and of an amorphous stripe solid with  
 phonon-type excitations. Using simple  scaling laws we then  determine the relationship 
 between the modulation length and the configurational entropy.}{}{}

\vspace*{10pt}
\keywords{Stripe glass, doped Mott insulator,  configurational entropy}
\textlineskip			
\vspace*{12pt}			

Competing interactions on different length scales are able to stabilize 
mesoscale phase separations and spatial inhomogeneities in a wide variety of 
systems. One important example in the context of correlated electron systems 
is the formation of stripes in doped Mott insulators, as found in transition 
metal oxides.\cite{CCJ93,JTra95}  The formation of a perfectly 
ordered array of stripes is however undermined by long range frustrating
 interactions.\cite{EK93} Very often, a long time dynamics emerges
similar to the relaxation seen
 in glasses.\cite{CBJ92}$^{-}$\cite{CH99} In particular the  slow, activated 
dynamics as obtained in NMR experiments  
 exhibits a striking universality, rather independent on 
the details of added impurities etc.\cite{JBC99,CH99} This suggests that glassiness in these 
systems is {\em self generated }and does not rely on the presence of 
quenched disorder, which will generally further stabilize a glassy state. 
 
Recently, two of us presented a theory for self generated randomness  caused 
by the competition between local   phase separation and a global long range 
interaction between domains of different charge\cite{SW00}. Using a replica 
approach developed for the investigation of structural
 glasses\cite{Mon95,MP991} and a numerical solution of the resulting many body problem it 
was demonstrated that a finite configurational entropy density, $S_{c}/V$, 
emerges at low temperatures. $S_{c}=k_{{\rm B}}\log {\cal N}_{{\rm ms}}$ is  
determined by the number of metastable states, ${\cal N}_{{\rm ms}}$, of the system.
 A finite $S_{c}/V$   implies that an exponential large 
(with the system size, $V$) number of metastable states exists. This is generally 
considered as a sign for the onset of glassy dynamics.\cite{KTW89} Given this enormous 
amount of metastable states, separated by high barriers, it becomes 
impossible for the system to explore the entire phase space and it  
gets trapped into metastable states, loosing the contribution $S_{c}$ of its entropy. 

The model studied in Ref.\cite{SW00}, which  characterizes a uniformly frustrated 
system with competition on different length scales, is given by\cite{EK93}:  
\begin{eqnarray} 
{\cal H} =\frac{1}{2}\int d^{3}x\left\{ r\varphi ({\bf x)}^{2}+\left( 
\nabla \varphi ({\bf x)}\right) ^{2}+\frac{u}{2}\varphi ({\bf x)}%
^{4}\right\}  
+\frac{Q}{2}\int d^{3}x d^{3}x^{\prime }\frac{\varphi ({\bf x)}\varphi 
({\bf x}^{\prime })}{\left| {\bf x-x}^{\prime }\right| }.  \label{ham11} 
\end{eqnarray} 
Here, $\varphi ({\bf x})$ characterizes charge degrees of freedom, with $%
\varphi ({\bf x})>0$ in \ a hole-rich region, $\varphi ({\bf x})<0$ in a 
hole poor region, and $\varphi ({\bf x})=0$ if the local density equals the 
averaged one. If \ $r<0$ the system tends to phase separate since we have to 
guarantee charge neutrality $\left\langle \varphi \right\rangle =0$. The 
coupling constant, $Q$, is a measure for the frustration between this short 
range coupling and the long range Coulomb interaction. An RPA-type  analysis 
of the correlation function for a system with Hamiltonian, Eq.\ref{ham11}, 
yields modulated structures, 
characterized by a modulation length, $l_{m}$, and an overall correlation 
length, $\xi $.\cite{NRK99}  Glassiness occurs  for $\xi \gg l_{m}$.\cite{SW00}
In this regime  the 
modulation wave number is given by $q_{m}\equiv \frac{2\pi }{l_{m}}\sim Q^{1/4}$. 
 
As  discussed in Ref.\cite{SW00} $S_{c}$ can be determined, following Ref. 
\cite{Mon95}\cite{MP991}, from the $m$-times replicated  partition function  
\begin{eqnarray} 
Z(m) =\lim_{g\rightarrow 0}\int D^{m}\varphi \exp \left( -\sum_{a=1}^{m}%
{\cal H}\left[ \varphi ^{a}\right] /T  -\frac{g}{2m}\sum_{a,b=1}^{m}\int d^{d}x\varphi ^{a}
({\bf x})\varphi ^{b}({\bf x})\right)
\end{eqnarray} 
via  
\begin{equation} 
S_{c}=-\left. \frac{\partial  \log Z(m)/m}{\partial m}\right| _{m=1}. 
\label{conf1} 
\end{equation} 
Here, $g$ is an infinitesimally small symmetry breaking field which causes a 
weak coupling between different replicas. Similar to the case of a small  
magnetic field at a ferromagnetic phase transition, $g$ drives a glassy 
system into a state with finite configurational entropy even in the limit
 $ g\rightarrow 0$. 
 
In this paper we demonstrate that a very transparent insight into the 
physical role of the configurational entropy can be obtained following the 
concept of replica bound states.\cite{MP991} Let $\Phi \left( {\bf r}\right) =%
\frac{1}{m}\sum_{a=1}^{m}\varphi ^{a}({\bf x})$ be the averaged 
configuration in replica space such that  
\begin{equation} 
\varphi ^{a}({\bf x})=\Phi \left( {\bf r}\right) +\psi ^{a}({\bf x}) 
\label{harans} 
\end{equation} 
with additional condition $\sum_{a=1}^{m}\psi ^{a}({\bf x})=0$. In the 
liquid state, the different replica configurations are completely 
independent and there is no inter-replica correlation in the system. On the 
other hand, within a glassy state (which we assume to exist based on our 
earlier  results) one expects that the various replicas of the 
system are tightly bound. Then, the deviations, $\psi ^{a}$, 
from the averaged replica configurations are  small compared to $\Phi $, 
allowing a harmonic approximation.\cite{MP991} Inserting Eq.\ref{harans} into the 
Hamiltonian of the system yields an effective propagator of the  $(m-1)$ replica fluctuations
\begin{equation} 
{\cal G}_{{\bf x,x}^{\prime }}^{-1}\left[ \Phi \right] =\left( r-\nabla 
^{2}+3u\Phi ^{2}\left( {\bf x}\right) \right) \delta _{{\bf x,x}^{\prime }}%
{\bf \ }+Q/\left| {\bf x-x}^{\prime }\right| \, , 
\end{equation} 
which depends on  $\Phi $. The Gaussian integral 
with respect to the $\psi^a$ can be performed and one can 
express $Z\left( m\right) $ in terms of $\Phi $ only: $
Z\left( m\right) =\int D\Phi e^{-\frac{m}{T}{\cal H}\left[ \Phi \right] -%
\frac{m-1}{2}{\rm tr}\ln {\cal G}^{-1}\left[ \Phi \right] }$. 
It is now straightforward to determine the configurational entropy using Eq.%
\ref{conf1}. It follows  
\begin{equation} S_{c}=S_{{\rm liq.}}+\left( \frac{1}{2}\left\langle {\rm tr}\ln {\cal G}\left[ \Phi %
\right] \right\rangle -V \right). 
\end{equation} 
The first term is the entropy of a  stripe-liquid.
The second term is the 
entropy of harmonic excitations relative to the slowly varying field $%
\Phi $, where  $N$ is the system size.
 We interpret $S_{{\rm sol.}}=-\frac{1}{2}\left\langle {\rm tr}\ln  
{\cal G}\left[ \Phi \right] \right\rangle +V$ as the entropy of an amorphous 
stripe solid (the stripe glass) with phononic  excitations.
As expected $S_c$ is the difference 
of the liquid entropy and the entropy of the amorphous solid.\cite{MP991} 
 
In order to estimate for the value of $S_{c}$ we proceed as follows. 
In the liquid state we assume that the entropy is determined by an 
effective Gaussian behavior, i.e. $S_{{\rm liq.}}=- \frac{1}{2}\left\langle  
{\rm tr}\ln {\cal G}_{0}\right\rangle +V$ with liquid state correlation 
function ${\cal G}_{0}$, where $\Phi ^{2}\left( {\bf r}\right) \simeq 
\left\langle \Phi ^{2}\right\rangle $. The amorphous solid on the other hand 
can be imagined as a liquid with additional confining potential  
\begin{equation} 
W\left( {\bf x}\right) =3u\left[ \Phi ^{2}\left( {\bf r}\right) 
-\left\langle \Phi ^{2}\right\rangle \right] . 
\end{equation} 
This confining potential reduces the degrees of freedom relative to the 
liquid state and therefore determines the amount of entropy lost in a glassy 
state, i.e. the configurational entropy. Using
 ${\cal G}\left[ \Phi \right]^{-1}={\cal G}_{0}^{-1}+W$ we find  
\begin{eqnarray} 
S_{c} &=&-\frac{1}{2}\left\langle {\rm tr}\ln \left( 1+W{\cal G}_{0}\right) 
\right\rangle   \nonumber \\ 
&\simeq &\frac{1}{4}{\rm \ }\int d^{3}rd^{3}r^{\prime }\left\langle W\left(  
{\bf x}\right) W\left( {\bf x}^{\prime }\right) \right\rangle {\cal G}%
_{0}\left( {\bf x-x}^{\prime }\right) ^{2},  \label{cc2} 
\end{eqnarray} 
where we restrict ourselves for the moment to the lowest order 
contribution. Higher order contributions will be discussed below.
A simple mean field analysis of Eq.\ref{ham11} suggests that the most important 
configurations of $W\left( {\bf x}\right) $ behave like $W\left( {\bf x}%
\right) =q_{m}^{2}f\left( q_{m}x\right) $ with dimensionless function $%
f $. Thus, the correlation function ${\cal R}\left( {\bf x-x}%
^{\prime }\right) =\left\langle W\left( {\bf x}\right) W\left( {\bf x}%
^{\prime }\right) \right\rangle $ behaves like ${\cal R}\left( x\right) 
=q_{m}^{4}\widetilde{f}\left( q_{m}x\right) $. On the other hand we know 
that the liquid state correlation function has a scaling behavior ${\cal G}%
_{0}\left( x\right) =q_{m}g\left( q_{m}x\right) $ if $q_m \gg \xi^{-1}$. Both, $\widetilde{f}$ and  
$g$ are dimensionless functions as well. Inserting these results into Eq.\ref 
{cc2} gives  
\begin{equation} 
S_{c}=CV(a q_{m})^{3}\sim VQ^{3/4}  \label{fin} 
\end{equation} 
with dimensionless constant $C$ and length scale $a$ of the order of the
 inter-atomic distance. In agreement with the numerical results of 
Ref.\cite{SW00} we find that the configurational entropy vanishes for $%
Q\rightarrow 0$. The larger the modulation length $l_{m}\sim q_{m}^{-1}$,
the smaller is the number of metastable states 
for obvious geometrical reasons. This gives a clear physical picture for the 
physical origin of the available number of metastable states within a 
stripe glass.
Note, our result, Eq.\ref{fin} is not altered if one goes beyond the leading 
term of the expansion of $\ln \left( 1+V{\cal G}_{0}\right) $. A 
diagrammatic expansion immediately shows that a contribution of order $n$ 
contains $n$ integrations like $\int d^{3}x$ $...$ as well as a term 
which behaves as ${\cal R}^{n/2}{\cal G}_{0}^{n}$. This leads to a 
dependence on the modulation length as $\left( q_{0}^{4}\right) 
^{n/2}q_{0}^{n}q_{0}^{-3\left( n-1\right) }=q_{0}^{3}$ where we took into 
account that one of the $x$- integrations gives an overall volume, $V$. 
Thus, Eq.\ref{fin} holds to arbitrary order of the expansion. 
 
In summary, we obtained an estimate for the  
configurational entropy of a stripe glass.
 $S_{\rm c}$ is the difference between the 
entropy of the liquid and the glass with harmonic 
phonon-like excitations. We analyzed this result by assuming that the 
amorphous solid can be generated ''perturbatively'' from the liquid state 
due to an additional confining potential, which reduces the degrees of 
freedom of the solid. Using  scaling laws for the relevant correlation 
functions we then found a  
relationship between $S_c$ and the stripe modulation length. 
 
\nonumsection{Acknowledgements}
\noindent
The work was supported by the Institute for Complex Adaptive Matter, by 
NSF (P.G.W.), Grant No. ChE-9530680 and by  Ames Laboratory (J.S.+H.W.Jr.),
operated for the U.S. Department of Energy by Iowa State  University 
under Contract  No. W-7405-Eng-82.


\begin{thebibliography}{100}
\bibitem{CCJ93}  J. H. Cho, F. C. Chou, and D. C. Johnston, Phys. Rev. Lett.  
{\bf 70}, 222 (1993). 
 
\bibitem{JTra95}  J. M. Tranquada, B. J. Sternlieb, J. D. Axe, Y. Nakamura, 
and S. Uchida, Nature {\bf 375}, 561 (1995).  
 
\bibitem{EK93}  V. J. Emery and S. A. Kivelson, Physica {\bf 209C}, 597 
(1993).  
 
\bibitem{CBJ92}  J. H. Cho, F. Borsa, D. C. Johnston, and D. R. Torgeson, 
Phys. Rev. B{\bf \ 46}, 3179 (1992). 
 
\bibitem{LC97}  S. H. Lee and S. W. Cheong, Phys. Rev. Lett.{\bf \ 79}, 2514 
(1997). 
 
\bibitem{TUI99}  J. M. Tranquada, N. Ichikawa, and S. Uchida, Phys. Rev.  
{\bf 59}, 14712 (1999). 
 
\bibitem{JBC99}  M.-H. Julien, F. Borsa, P. Carretta, M. Horvatic, C. 
Berthier, and C. T. Lin, Phys. Rev. Lett. {\bf 83}, 604 (1999). 
 
\bibitem{HS99}  A. W. Hunt, P. M. Singer, K. R. Thurber, and T. Imai, Phys. 
Rev. Lett. {\bf 82}, 4300 (1999). 
 
\bibitem{CH99}  N. J. Curro, P. C. Hammel, B. J. Suh, M. H\"{u}cker, B. B%
\"{u}chner, U. Ammerahl, and A. Revcolervschi, Phys. Rev. Lett. 
{\bf 85},642 (2000).
 
\bibitem{SW00}  J. Schmalian and P. G. Wolynes, Phys. Rev. Lett 
{\bf 85}, 836 (2000). 
 
\bibitem{Mon95}  R. Monnason, Phys. Rev. Lett. {\bf 75}, 2875 (1995). 
 
\bibitem{MP991}  M. Mezard and G. Parisi, Phys. Rev. Lett. {\bf 82}, 747 
(1999).  
 
\bibitem{NRK99}  Z. Nussinov, J. Rudnick, S. A. Kivelson, L. N. Chayes, 
Phys. Rev. Lett. {\bf 83}, 472 (1999). 
 
\bibitem{KTW89} T. R. Kirkpatrick and P. G. Wolynes, Phys. Rev. A {\bf 35}, 3072 (1987),
 T. R. Kirkpatrick and D. Thirumalai, and P. G. Wolynes, 
Phys. Rev. A {\bf 40}, 1045 (1989); T. R. Kirkpatrick and P. G. Wolynes, 
Phys. Rev. B {\bf 36}, 8552 (1987);T. R. Kirkpatrick and D. Thirumalai, 
Phys. Rev. Lett. {\bf 58}, 2091 (1987).  


\end{thebibliography}
\end{document}